\documentclass[prb,amsmath,amssymb,showkeys]{revtex4}
\usepackage{bm}
\def\be{\begin{equation}}
\def\ee{\end{equation}}
\def\nab{\bm{\nabla}}
\def\qc{\frac{q}{c}}
\def\ene{\tilde{n}}
\begin{document}
\title{Extension of the de Broglie-Bohm theory to the
Ginzburg-Landau equation}
\author{Jorge Berger}
\affiliation{Physics Unit, Ort Braude College, P. O. Box 78,
21982 Karmiel, Israel and \\
Department of Physics, Technion, 32000 Haifa, Israel}
\email{phr76jb@tx.technion.ac.il}
\begin{abstract}
The de Broglie-Bohm approach permits to assign well defined trajectories
to particles that obey the Schroedinger equation. We extend this
approach to electron pairs in a superconductor. In the stationary regime
this extension is completely natural; in the general case additional 
postulates are required. This approach gives enlightening
views for the absence of Hall effect in the stationary regime and for the
formation of permanent currents.
\end{abstract}
\maketitle
\section{INTRODUCTION}
If the evolution of a particle obeys the Schroedinger equation, then
the de Broglie-Bohm quantum theory (dBB)\cite{Brog,1,BH} provides
a deterministic description of its motion by assuming that, besides
the classical forces, an additional ``quantum force"
$\nab (\hbar^2\nabla^2|\psi|/2m|\psi|)$ acts on the particle. The dBB
theory remains applicable in the presence of magnetic fields.\cite{2}
Variations and extensions of dBB have recently been considered.\cite{Floyd,John}

In this article we want to extend dBB to the motion of Cooper pairs in a 
superconductor, which obey the Ginzburg-Landau equation\cite{GL,Sch1}
\begin{equation}
\frac{1}{2m}\left(-i\hbar\nab-\qc{\bf A}\right)^2\psi+\alpha\psi+\beta|\psi|^2\psi=
-\gamma\hbar\frac{\partial\psi}{\partial t} \;.
\label{GL}
\ee
Here we follow the notation of Ref.~\onlinecite{Sch1}, that makes Eq.~(\ref{GL}) look 
similar to the Schroedinger equation: $\psi$ is the ``order parameter", $t$ is the 
time, $q$ and $m$ are the charge 
and the mass of a pair, $\hbar$ and $c$ are Plank's constant and the speed of light, 
$\alpha$, $\beta$ and $\gamma$ are material constants,
and ${\bf A}$ is the vector potential of the electromagnetic field. We have chosen a 
gauge such that the scalar potential vanishes.

In this letter we will interpret the order parameter $\psi=|\psi|e^{i\varphi}$ as 
the vawefunction for Cooper pairs.
As usual, $|\psi|^2$ will have the meaning of density of particles, $\hbar\nab\varphi$
will be the canonic momentum, and ${\bf v}=(\hbar\nab\varphi-\qc{\bf A})/m$,
the velocity of the pairs. However, the equation of motion will not be based on the
usual Hamilton--Jacobi formalism, but just on Newton's second law. The price we
will have to pay for this approach is some basic knowledge of fluid mechanics
(e.g. Ref.~\onlinecite{fluid}).

\section{STATIONARY SITUATIONS}
In this case the right hand side of Eq.~(\ref{GL}) vanishes. The remaining equation
differs qualitatively from the Schroedinger equation, since it is nonlinear. We shall
see that, nevertheless, the dBB theory can be smoothly extended.
 
\subsection{Equation of Motion}
For a stationary situation, conservation of particles gives
$\nab\cdot(|\psi|^2{\bf v})=0$, which leads to
\be
2(\hbar\nab\varphi-\qc{\bf A})\cdot\nab|\psi|+|\psi|(\hbar\nabla^2\varphi-
\qc\nab\cdot{\bf A})=0 \;.
\label{cont}
\ee
Expanding Eq. (\ref{GL}), using Eq. (\ref{cont}), and defining 
\be
-Q_{\rm stat}\equiv\frac{\hbar^2}{2m}\frac{\nabla^2|\psi|}{|\psi|}-\beta|\psi|^2  \;,
\label{Q}
\ee
we arrive at
\be
-Q_{\rm stat}=\frac{1}{2m}(\hbar\nab\varphi-\qc{\bf A})^2+\alpha \;.
\label{541}
\ee

Taking the gradient at both sides of Eq.~(\ref{541}) and noting that in the
stationary regime ${\bf a}=({\bf v}\cdot\nab){\bf v}$ is the convective time derivative
of the velocity, and hence the acceleration of a pair, we obtain
\be
-\nab Q_{\rm stat}+q\frac{\bf v}{c}\times{\bf B}=m{\bf a} \;,
\label{newton}
\ee
with ${\bf B}=\nab\times{\bf A}$. This means that, in addition to the magnetic
force, the quantum force $-\nab Q_{\rm stat}$ acts on every pair.
As a matter of principle, $-\nab Q_{\rm stat}$ is not only the quantum force,
but it also includes the force exerted by the lattice. Note that the only
physical input required for Eq.~(\ref{newton}) is Eq.~(\ref{GL}) (with vanishing
r.h.s.) and the interpretations made in the Introduction; the material
parameters, the applied field or the geometry of the superconductor have
no influence.

\subsection{The Hall Effect}
We shall see that Eq. (\ref{newton}) provides an intuitive way to explain the 
absence of Hall effect in a superconductor in a stationary state. 
The Hall effect was
discovered in 1879, well before quantum effects were suspected to exist.
Historically, it turned out that an apparently trivial classical explanation 
could be given to the effect (except for the existence of positive carriers).
Only when the Hall voltage turned out not to be a linear function of the 
magnetic field,\cite{QHE} the effect was called ``quantum". 

In the case of
a superconductor carrying a stationary current perpendicular to a magnetic 
field that penetrates the sample, there is clearly a Lorentz force that
acts on the flowing electrons; therefore, it seems obvious to anticipate
the presence of an electric force that balances it, implying a Hall 
voltage.\cite{Lewis} 

However, early experiments found that the Hall voltage
drops to zero when a sample becomes superconducting.\cite{0,Lew0}
Moreover, if there were an electric field in the superconductor,
it would produce normal currents and dissipation. At least in the case of 
permanent currents, this is not what is observed. 

The resolution of the 
paradox is provided by Eq.~(\ref{newton}): the Lorenz force is balanced by
the quantum force $-\nab Q_{\rm stat}$ in order to complete the total force required
to keep the pairs on their trajectories, and no electric force is required.

Later experiments did find a Hall voltage in superconductors (e.g. 
Ref.~\onlinecite{Kop}). However, in these cases the voltage is due to the motion of
vortices and cannot be described as a stationary situation.

\subsection{An Example}
As an illusration, let us examine a situation with constant field and
cylindric symmetry. Let the field be ${\bf B}=B_0\hat{z}$ and the vector 
potential, ${\bf A}=\frac{1}{2}B_0 r\hat{\theta}$, where $r$, $\theta$ and $z$
are cylindric coordinates. Cylindric symmetry implies that $\varphi$ is a linear
function of $\theta$. Together with single-valuedness, this enables us to write
$\varphi=-n\theta$, where $n$ is an integer.
It follows that ${\bf v}=(1/m)(|q|B_0 r/2c-n\hbar/r)\hat{\theta}$ and therefore
the sum of the Lorentz and the centrifugal force is
$${\bf F}_{\rm Lorentz}+{\bf F}_{\rm centrifugal}=\frac{1}{mr}
\left(\frac{n^2\hbar^2}{r^2}-\frac{q^2B_0^2 r^2}{4c^2}\right)\hat{r}\;.$$
This force points away from or towards the $z$-axis, depending on whether $r$ is
smaller or greater than $r_B=(2nc\hbar/|q|B_0)^{1/2}$.
On the other hand, using Eq.~(\ref{541}) we find 
$$-Q_{\rm stat}=\frac{1}{2m}\left(\frac{n\hbar}{r}-\frac{|q|B_0 r}{2c}\right)^2+\alpha\;.$$
(The quantum potential is maximal at $r=r_B$.) We can now immediately verify that
$-\nab Q_{\rm stat}=-{\bf F}_{\rm Lorentz}-{\bf F}_{\rm centrifugal}$.

\section{GENERAL CASE}
Now Eq.~(\ref{GL}) is not only nonlinear; it is also nonunitary.
Accordingly, the number of Cooper pairs is not conserved and any model that is
limited to a fixed number of particles cannot describe the physical situation.
We can still envision a deterministic evolution in the spirit of Refs. \onlinecite{Sp}
and \onlinecite{crea} (``beables"). The objects of the model are the order parameter 
$\psi$, which acts as a field, and the density of particles. 
We still postulate that particles move with 
velocity $(\hbar\nab\varphi-\qc{\bf A})/m$. The order parameter, the density and
the velocity depend on position and on time. The evolution of the system will consist
of two processes: motion of pairs and creation (or destruction) of pairs. In order
to have a consistent model, we require that if the density of pairs is initially
$|\psi|^2$, it will remain given by $|\psi|^2$ as the system evolves.

\subsection{Pair Trajectories}
We expand Eq.~(\ref{GL}) and factor out $e^{i\varphi}$. Equating the imaginary
parts we obtain
\be
\nab\cdot(|\psi|^2{\bf v})=2\gamma|\psi|^2\frac{\partial\varphi}
{\partial t}
\label{Im}
\ee
and, from the real parts
\be
-Q_{\rm stat}=\alpha+\frac{m}{2}v^2+\frac{\gamma\hbar}{2|\psi|^2}
\frac{\partial|\psi|^2}{\partial t}\;.
\label{Re}
\ee

We define now
\be
-Q_{\rm dyn}\equiv -\frac{\hbar}{2|\psi|^2}\left(\gamma\frac{\partial|\psi|^2}
{\partial t}-\frac{1}{\gamma}\nab\cdot(|\psi|^2{\bf v})\right)   \;.
\label{Qdyn}
\ee
Taking the gradient at both sides of Eq.~(\ref{Re}), using Eq.~(\ref{Im}), 
noting that the acceleration is now 
${\bf a}=({\bf v}\cdot\nab){\bf v}+\partial{\bf v}/\partial t$ and the
electric field is ${\bf E}=-(1/c)\partial{\bf A}/\partial t$ gives
\be
-\nab (Q_{\rm stat}+Q_{\rm dyn})+q\left({\bf E}+\frac{\bf v}{c}\times{\bf B}\right)=
m{\bf a} \;,
\label{newtondyn}
\ee
so that the motion of the pairs can still be interpreted as having Newtonian
trajectories, provided that the quantum force $-\nab (Q_{\rm stat}+Q_{\rm dyn})$
is added to the electromagnetic force.
Note that now the quantum force is not a function of $|\psi|$ only, but is still
a functional of $\psi$.

In regions where $|\psi|$ is nearly constant and uniform, 
$-\nab Q_{\rm dyn}\approx\frac{\hbar}{2\gamma}\nabla^2{\bf v}$. In this case
$-\nab Q_{\rm dyn}$ behaves as a viscous force, with viscosity coefficient
$\frac{\hbar}{2\gamma}|\psi|^2$.

\subsection{Creation and Annihilation of Particles}
We build a model for formation and destruction of pairs as follows. We define
the depairing potential as
\be
Q_{\rm dep}\equiv 2\left(\gamma+\frac{1}{\gamma}\right)\left(Q_{\rm stat}+
 \alpha+\frac{m}{2}v^2\right)+2\gamma Q_{\rm dyn}\;.
\label{Qopt}
\ee
We postulate that at intervals of time $\tau$ the system checks itself.
(For typical non-stationary processes in superconductors, characteristic times
are of the order of $10^{-13}$s.)
At the end of each interval we divide the sample into ``depairing cells" such
that the volume integral in each cell obeys 
$|\int Q_{\rm dep}|\psi|^2dV|=\hbar/\tau$. Since the integral of 
$Q_{\rm dep}|\psi|^2$ over the entire sample is not necessarily integer, some
marginal region of the sample will in general be left out of the cells.
In order to have a well defined division we need some additional criterion;
for instance, we may require that the average boundary area of the cells be
as small as possible. This criterion would discourage elongated cells or 
merging of cells with positive and negative $Q_{\rm dep}$; also, the regions
left out of the cells will be those where $|Q_{\rm dep}||\psi|^2$ is small.
Finally, we postulate that if $\int Q_{\rm dep}|\psi|^2dV$ is negative, then
a new Cooper pair forms at the center of the cell and, if it is positive, 
the pair closest to the center is destroyed. 

We have to prove that the evolution of the pair density predicted by this model
is the same as that predicted by Eq.~(\ref{GL}). The pair density changes due to
two processes: flow of pairs and net pair creation. The increase of density per
unit time do to flow is $-\nab\cdot(|\psi|^2{\bf v})$. The increase due to
creation equals the net number of negative cells per unit volume,
$-Q_{\rm dep}|\psi|^2\tau/\hbar$, multiplied by the frequency $1/\tau$.
(We have assumed that the cells and $\tau$ are sufficiently small to be treated
as a continuum, with $Q_{\rm dep}|\psi|^2$ uniform in the analyzed region.)
Substituting Eqs. (\ref{Re}) and (\ref{Qdyn}) into Eq.~(\ref{Qopt}) we obtain 
that the total rate of 
density increase is $\partial |\psi|^2/\partial t$, as required.

For fast processes we expect that typical depairing
cells will have microscopic sizes, but if the particle density is small or
a process is almost stationary, these cells can be large. We could build an 
alternative model by postulating the existence of depairing cells with fixed
positions and sizes, which destroy or create a pair every time that the
integral $\int\!\int Q_{\rm dep}|\psi|^2dV\,dt$ in the cell changes by $\pm\hbar$.

\subsection{Example}
We consider ring of radius $R$, sufficiently thin to be treated as one-dimensional,
threaded by a magnetic flux $\Phi$. For negative times the ring is in the normal
state, but at $t=0$ it is instantaneously cooled and becomes superconducting.
In Eq.~(\ref{GL}) cooling below the critical temperature is implemented by changing
the value of $\alpha$ from positive to negative. It is well known that if $\Phi$
is not an integer multiple of the quantum of flux $\Phi_0=ch/|q|$, then a 
permanent current flows around the ring. A classical-minded question naturally
arises: since there are no electromagnetic fields in the ring, how do the charges
know to start moving when the temperature is lowered?

In order to answer this question we first have to calculate $\psi$. Let us
restrict our attention to the period during which $|\psi|$ is so small that the 
nonlinear term in Eq.~(\ref{GL}) can be neglected. For a thin
ring Eq.~(\ref{GL}) takes the one-dimensional form
$$\frac{\hbar^2}{2mR^2}\left(i\frac{\partial}{\partial\theta}-\frac{\Phi}{\Phi_0}
\right)^2\psi+\alpha\psi=-\hbar\gamma\frac{\partial\psi}{\partial t}\;.$$
We write $\psi=\sum_{n=-\infty}^{\infty}T_n(t)e^{-ni\theta}$ and linearity permits
to treat each harmonic separately. We obtain 
$$T_n(t)=T_n(0)e^{\lambda_n t}\;,\;\;\lambda_n=-\frac{\alpha}{\hbar\gamma}
-\frac{\hbar(n-\Phi/\Phi_0)^2}{2m\gamma R^2\;.}$$
Since the ring was initially in the normal state, the initial values $T_n(0)$
are very small. (They don't vanish due to thermal fluctuations.\cite{Sch2})

For simplicity, let us consider the case in which $\lambda_n>0$ for just one
value of $n$ (the closest to $\Phi/\Phi_0$), which will be denoted by $\ene$.
This means that $T_{\ene}$ will increase with time, whereas $T_n$ will remain 
negligible for $n\neq\ene$. As a consequence, $|\psi|$ will be independent
of position.

After $\psi$ is known, we can evaluate the potentials. We obtain $Q_{\rm stat}=0$,
$Q_{\rm dyn}=\hbar\gamma\lambda_{\ene}$ and $Q_{\rm dep}=-2\hbar\lambda_{\ene}<0$.
Since these potentials are independent of position, it follows that the pairs are not
accelerated. Current arises due to their formation, with velocity
$(\hbar/mR)(\Phi/\Phi_0-\ene)\hat\theta$.

\section{CONCLUSION}
We have extended the de Broglie-Bohm theory, which was built for particles that
obey the Schroedinger equation, to the case of electron pairs in a superconductor,
which obey a more complex (nonlinear and nonunitary) equation. For the stationary
regime this extension is completely natural; in the general case, in which the
number of pairs is not conserved, additional postulates are required. 

\begin{acknowledgments}
I am grateful to Mark Israelit and Jacob Rubinstein for constructive comments
to the first draft of this letter and to Lev Vaidman and Oscar Chavoya for
bringing my attention to Refs. \onlinecite{Sp} and \onlinecite{crea}. The example
in the dynamic case was inspired by an article by Jorge Hirsch in cond-mat.
\end{acknowledgments}

\end{document}